\documentclass{jfm}
\usepackage{bm}
\def\dt{{\rm d}t}
\def\d{{\rm d}}
\def\dk{{\rm d}k}

\def\x{{\bm x}}
\def\k{{\bm k}}
\def\etal{{\it et al. }}

\title[Remarks on the KLB theory of 2D turbulence]
{Remarks on the KLB theory of two-dimensional turbulence}
\author[C.V. Tran and T.G. Shepherd]
{C\ls H\ls U\ls O\ls N\ls G\ns V.\ns T\ls R\ls A\ls N$^1$\ns \and T\ls H\ls 
E\ls O\ls D\ls O\ls R\ls E\ns G.\ns S\ls H\ls E\ls P\ls H\ls E\ls R\ls D$^2$}
\affiliation{$^1$Mathematics Institute, University of Warwick, Coventry 
CV4 7AL, UK \\[\affilskip]
$^2$Department of Physics, University of Toronto, 60 St. George Street,
Toronto, Ont., Canada M5S 1A7 \\[\affilskip]}

\begin{document}

\maketitle

\begin{abstract}

We study the inverse energy transfer in forced two-dimensional (2D) 
Navier--Stokes turbulence in a doubly periodic domain. It is shown that 
an inverse energy cascade that carries a nonzero fraction of the 
injected energy to the large scales via a power-law energy spectrum 
$\propto k^{-\alpha}$ requires that $\alpha\ge5/3$. This result is 
consistent with the classical theory of 2D turbulence that predicts a 
$k^{-5/3}$ inverse-cascading range, thus providing for the first time a 
rigorous basis for this important feature of the theory. We derive bounds 
for the Kolmogorov constant $C$ in the classical energy spectrum 
$E(k)=C\epsilon^{2/3}k^{-5/3}$, where $\epsilon$ is the energy injection 
rate. Issues related to Kraichnan's conjecture of energy condensation 
and to power-law spectra as the quasi-steady dynamics become steady are 
discussed.    

\end{abstract}

\section{Introduction}

It is well known that the advective nonlinearities of two-dimensional 
(2D) Navier--Stokes (NS) turbulence predominantly transfer energy to low 
wavenumbers (large scales) and enstrophy to high wavenumbers (small scales). 
The extent of this preferential transfer has been a subject of intense 
research since \cite{Fjortoft53} first noticed this interesting property 
of 2D turbulence. In the late 1960s \cite{Kraichnan67} postulated that for 
unbounded fluids in the limit of high Reynolds number, this preferential 
transfer achieves an extreme limit, by transferring virtually all energy to 
ever-lower wavenumbers (inverse energy cascade) and virtually all enstrophy 
to a high wavenumber $k_\nu\gg s$ (direct enstrophy cascade) if the turbulence 
is driven by sources at intermediate wavenumbers around $s$. The transfer of 
energy and enstrophy in this extreme manner is known as the dual cascade. 
The energy is predicted to cascade via a $k^{-5/3}$ energy inertial range, 
and the enstrophy is predicted to cascade via a $k^{-3}$ enstrophy inertial 
range. The dissipation wavenumber $k_\nu$ determines the region where the 
enstrophy gets dissipated. The resulting dynamics is quasi-steady since 
the inverse energy cascade presumably proceeds indefinitely in time 
toward wavenumber zero. This theory was later advanced by \cite{Leith68}, 
\cite{Batchelor69}, and \cite{Kraichnan71} and has become a classical 
theory of 2D turbulence, known as the KLB theory.

For a fluid confined to a doubly periodic domain, the inverse energy
cascade is halted at the lowest wavenumber $k_0$ corresponding to the 
integral length scale of the system. This arrest of the inverse 
cascade may be met with complex responsive adjustments of the $k^{-5/3}$
range. In any case a rise of the total energy in the available energy
range necessarily occurs. According to \cite{Kraichnan67}, such a rise 
occurs only at $k_0$, resulting in what can be termed an `energy 
condensate' that singlehandedly carries most of the total kinetic 
energy.\footnote{\cite{Kraichnan67} also predicts that as the energy 
accumulates at $k_0$, the $-5/3$ spectrum is modified toward absolute 
equilibrium of the form $E(k)\propto k/(\beta k^2+\alpha)$, where $\beta$ 
and $\alpha$ are constant (see Kraichnan 1967, p.~1423a). Note that 
Kraichnan's conjecture would be trivially extended to an energy condensate 
in the vicinity of $k_0$, not just at $k_0$. In this work we address this 
more general scenario.} Assuming that the energy spectrum eventually 
becomes statistically steady, with a persistent inverse cascade up to 
$k_0$, the energy condensate 
not only would carry virtually all the system energy but also would 
account for virtually all the energy dissipation. 
This means that Kraichnan's energy condensate would also be an enstrophy 
condensate, although the latter would be of a lesser degree. For a 
sufficiently wide energy inertial range, the enstrophy dissipation by the 
energy condensate would be negligible. This possibility allows for the 
enstrophy dynamics to be virtually unaffected by the condensate and for 
the direct enstrophy cascade to remain intact in Kraichnan's picture. 

Numerical results concerning the Kraichnan condensate are inconclusive
and, in fact, controversial. Smith \& Yakhot (1993, 1994) argue for a 
positive answer to the Kraichnan conjecture. On the contrary \cite{Borue94} 
and \cite{TB04} observe that after the arrest of the inverse energy 
cascade, growth of energy and enstrophy occurs throughout the energy 
inertial range. More precisely they find that as the turbulence approaches 
a steady state, a $k^{-3}$ energy spectrum forms at large scales. Although 
this spectrum still means that most of the energy is concentrated at the
lowest wavenumbers, it is by no measure close to the Kraichnan picture. 
In particular a $k^{-3}$ energy spectrum means a $k^{-1}$ enstrophy 
spectrum, so that no accumulation of enstrophy occurs at the largest scales
and the dissipation of energy cannot occur primarily at or in the vicinity 
of $k_0$.

Theoretical studies have shown under a variety of assumptions that for 
a doubly periodic domain, the ratio of enstrophy dissipation to energy
dissipation equals $s^2$---the ratio of enstrophy injection to energy
injection (Constantin \etal 1994; Eyink 1996, p. 110; Tran \& Shepherd 
2002; Kuksin 2004). These authors have concluded from this result that 
there can be no direct enstrophy cascade. However this inference assumes 
that all the enstrophy is actively involved in the enstrophy cascade, an 
assumption that is inconsistent with the Kraichnan conjecture. Indeed, 
several of these studies have explicitly or implicitly excluded a priori 
the Kraichnan conjecture in favour of an assumption of power-law spectra. 
There is thus a need to explicitly examine the extent to which the 
Kraichnan conjecture is consistent with analytical constraints on 2D 
turbulence.
 
In this paper we study several aspects of the KLB theory. In \S\,2 we 
introduce some necessary notation and derive a basic identity and inequality.  
In \S\,3 we give a general review of the KLB theory, including a discussion 
of the dual-cascade hypothesis and of the Kraichnan conjecture of energy 
condensation. In \S\,4 we derive an upper bound for the inverse energy flux 
and show that a nonzero steady flux requires that the inverse-cascading 
range be at least as steep as $k^{-5/3}$. More precisely, we show that an 
energy-transfer range shallower than $k^{-5/3}$ is unable to support a 
constant inverse energy flux. This result provides for the first time a 
rigorous basis for the Kolmogorov--Kraichnan $k^{-5/3}$ inertial range. 
We also derive bounds for the Kolmogorov constant $C$ in the classical
energy spectrum $E(k)=C\epsilon^{2/3}k^{-5/3}$, where $\epsilon$ is the 
energy steady injection rate. In \S\,5 we derive a constraint on the 
enstrophy and palinstrophy distribution in the Kraichnan picture and 
contrast it with power-law scalings. Concluding remarks and discussion
are given in \S\,6. 
 
\section{Mathematical preliminaries} 

In the vorticity formulation the forced 2D NS equation governing the 
motion of an incompressible fluid confined to a doubly periodic domain 
$[0,L]\times[0,L]$ is
\begin{eqnarray}
\label{NS}
\partial_t\Delta\psi+J(\psi,\Delta\psi) &=& \nu\Delta^2\psi +f,
\end{eqnarray}
where $\psi(\bm x,t)$ is the streamfunction, 
$J(\theta,\vartheta)=\theta_x\vartheta_y-\theta_y\vartheta_x$,
$\nu$ the kinematic viscosity and $f(\bm x,t)$ the forcing. The 
velocity field $\bm v(\bm x,t)$ can be recovered from the streamfunction 
$\psi(\bm x,t)$ by $\bm v=(-\psi_y,\psi_x)$. The nonlinear term admits 
the identities
\begin{eqnarray}
\label{conservation}
\langle \phi J(\theta,\vartheta)\rangle &=& 
-\langle \theta J(\phi,\vartheta)\rangle =
-\langle \vartheta J(\theta,\phi)\rangle,
\end{eqnarray}
where $\langle\,\cdot\,\rangle$ denotes a spatial average. As a consequence
we have the twin constraints
\begin{eqnarray} 
\langle\psi J(\psi,\Delta\psi)\rangle
&=& 0 = \langle\Delta\psi J(\psi,\Delta\psi)\rangle , 
\end{eqnarray}
so that the energy $E=\langle|\nabla\psi|^2\rangle/2$ and enstrophy 
$Z=\langle|\Delta\psi|^2\rangle/2$ are conserved by nonlinear transfer. 

We now derive a simple identity and an inequality, which are used in
\S\,4 to calculate the nonlinear triple-product term representing the 
inverse energy transfer. By straightforward calculation we have
\begin{eqnarray}
\Delta J(\theta,\vartheta) &=& J(\Delta\theta,\vartheta)+
J(\theta,\Delta\vartheta)+2J(\theta_x,\vartheta_x)+2J(\theta_y,\vartheta_y).
\end{eqnarray}
Hence
\begin{eqnarray}
\langle\Delta\theta J(\theta,\vartheta)\rangle &=& 
\langle\theta\Delta J(\theta,\vartheta)\rangle 
= \langle\theta J(\Delta\theta,\vartheta)\rangle
+2\langle\theta J(\theta_x,\vartheta_x)\rangle
+2\langle\theta J(\theta_y,\vartheta_y)\rangle \nonumber\\
&=& -\langle\Delta\theta J(\theta,\vartheta)\rangle
+2\langle\theta J(\theta_x,\vartheta_x)\rangle
+2\langle\theta J(\theta_y,\vartheta_y)\rangle.
\end{eqnarray}
It follows that
\begin{eqnarray}
\label{id}
\langle\Delta\theta J(\theta,\vartheta)\rangle &=& 
\langle\theta J(\theta_x,\vartheta_x)\rangle +
\langle\theta J(\theta_y,\vartheta_y)\rangle.
\end{eqnarray} 

For the present case of a doubly periodic domain of size $L \times L$, 
the Fourier representation of $\psi(\bm x)$ is
\begin{eqnarray}
\psi(\x)=\sum_{\k}\exp\{i\k\cdot\x\}\widehat\psi(\k).
\end{eqnarray}
Here $\k=k_0(n,m)$, where $k_0=2\pi/L$ is the lowest wavenumber and $n$ 
and $m$ are integers not simultaneously zero. For a given wavenumber $\ell$ 
let $\psi^<$ and $\psi^>$ denote, respectively, the components of 
$\psi$ spectrally supported by the disk $d=\{\k\,:\,k<\ell\}$ and its 
complement $D=\{\k\,:\,k\ge \ell\}$, i.e.
\begin{eqnarray}
\psi^<=\sum_{\k\in d}\exp\{i\k\cdot\x\}\widehat\psi(\k),
~~~~\psi^>=\sum_{\k\in D}\exp\{i\k\cdot\x\}\widehat\psi(\k).
\end{eqnarray}
The lower-wavenumber component $\psi^<$ satisfies 
\begin{eqnarray}
\label{supbound}
|\Delta\psi^<| &\le& \sum_{\k\in d}k^2|\widehat\psi(\k)| 
\le \left(\sum_{\k\in d}1\right)^{1/2}
\left(\sum_{\k\in d}k^4|\widehat\psi(\k)|^2\right)^{1/2}
= c\frac{\ell}{k_0}Z_<^{1/2},
\end{eqnarray}
where $c$ is an absolute constant of order unity and 
$Z_<=\langle|\Delta\psi^<|^2\rangle/2$ the large-scale enstrophy 
density associated with the wavenumbers $k<\ell$. In (\ref{supbound}) 
the Cauchy--Schwarz inequality is used in the second step, and the sum 
$\sum_{\k\in d}1\approx \ell^2/k_0^2$ represents the number of 
wavevectors in $d$.

\section{The KLB theory}

This section reviews the central features of the KLB theory: the 
dual-cascade hypothesis and the Kraichnan conjecture of energy 
condensation. The discussion deviates significantly from the original 
works that lead to the dual-cascade hypothesis, thereby providing a 
new look at the KLB theory. We also discuss some recent results, which, 
on the one hand, can modify KLB significantly and, on the other hand, 
cast doubt on some aspects of the theory.

\subsection{The preferential transfer}

We consider turbulence driven by sources localized around a wavenumber 
$s$ that supply an energy injection $\epsilon > 0$ and an enstrophy 
injection $s^2\epsilon$. Let $\epsilon(k)$ and $k^2\epsilon(k)$ be 
nonlinear redistributions of these injections in wavenumber space. Given 
arbitrary wavenumbers $r_1<s$ and $r_2>s$ we have
\begin{eqnarray}
\label{pref1}
\frac{1}{\epsilon}\int_{r_2}^\infty \epsilon(k)\,\dk
&\le& \frac{1}{r_2^2\epsilon}\int_{r_2}^\infty k^2\epsilon(k)\,\dk 
\le \frac{s^2\epsilon}{r_2^2\epsilon} = \frac{s^2}{r_2^2},\\
\label{pref2}
\frac{1}{s^2\epsilon}\int_0^{r_1} k^2\epsilon(k)\,\dk &\le& 
\frac{r_1^2}{s^2\epsilon}\int_0^{r_1}\epsilon(k)\,\dk \le 
\frac{r_1^2\epsilon}{s^2\epsilon} = \frac{r_1^2}{s^2},
\end{eqnarray}
where the second inequalities in (\ref{pref1}) and (\ref{pref2}) are,
respectively, due to the conservation of enstrophy and energy. It follows 
that the respective fractions of $\epsilon$ and of $s^2\epsilon$ that 
get transferred to $k\ge r_2>s$ and to $k\le r_1<s$ are bounded from 
above by $s^2/r_2^2$ and by $r_1^2/s^2$, respectively. These constraints 
correspond to 
the well-known prohibition of a significant direct (inverse) transfer 
of energy (enstrophy): no considerable fraction of $\epsilon$ 
($s^2\epsilon$) is allowed to get transferred to wavenumbers $k\gg s$ 
($k\ll s$). Hence when these injections spread out in wavenumber space,
most of the energy (enstrophy) gets transferred toward lower (higher) 
wavenumbers, a qualitative conclusion that is behind the dual-cascade 
hypothesis. This dynamical behaviour of energy and enstrophy is well
confirmed by numerical simulations reported in the literature. Note that 
the preferential transfer described above reflects the collective effects
of all admissible nonlinearly interacting triads. Detailed analyses 
of individual triads may not lead to the same conclusion with certainty
since there is a significant fraction of triads that may act unfavourably 
(see Merilees \& Warn 1975).

\subsection{The dual-cascade hypothesis and Kraichnan conjecture}

The preferential transfer of energy and enstrophy, together with the
scale-selective dissipation by molecular viscosity, constitutes the 
backbone of the dual-cascade hypothesis. Let us consider the dynamical 
scenario described in the preceding subsection, with an additional 
assumption that the fluid is initially at rest. When the forcing region
becomes unstable and the injected enstrophy gets transferred toward the 
high wavenumbers, palinstrophy $P=\langle|\nabla\Delta\psi|^2\rangle/2$
is built up and growth of the enstrophy dissipation $2\nu P$ ensues. 
Meanwhile growth of enstrophy is due solely to enstrophy injection and is 
suppressed by its own growing dissipation. If the growth of palinstrophy is 
sufficiently rapid, the enstrophy can quickly become steady, achieving a 
value sufficiently low that $2\nu Z\ll\epsilon$. This allows for a 
strong inverse energy cascade to be realized. By `strong' we mean that the 
inverse cascade carries virtually all the energy injection to the large 
scales; otherwise the inverse cascade is `weak'. At least a {\it transient} 
direct enstrophy cascade seems plausible since the palinstrophy is 
concentrated mainly at high wavenumbers, presumably at $k>k_\nu\gg s$ 
(the so-called dissipation range in the KLB theory). The realization of 
the classical $k^{-3}$ enstrophy 
cascade as a persistent rather than just transient phenomenon requires 
that the high concentration of palinstrophy around $k_\nu$ (which may 
be correctly termed a `palinstrophy condensate') remain stable, 
withstanding the enormous dissipation of palinstrophy in that region 
and not redistributing itself in wavenumber space. Note that if viscosity 
is replaced by a scale-independent dissipation, i.e. a mechanical 
friction, then both energy and enstrophy are dissipated at the same rate. 
The above argument for an inverse cascade fails to apply in this case. 
If the usual viscosity term $\nu\Delta^2\psi$ is replaced by an 
inverse viscosity represented by, for example, $\mu\psi$, then the 
dynamical behaviours of energy and enstrophy are reversed: energy behaves 
as enstrophy and vice versa. In this case the preferential transfer of 
energy and enstrophy gives rise to a predominant increase of the energy 
dissipation $\mu\langle|\psi|^2\rangle$. The energy can then quickly become 
steady, achieving a value sufficiently low that $2\mu E\ll s^2\epsilon$,
i.e. the enstrophy dissipation is much less than its injection rate. This 
allows for a direct enstrophy cascade to be realized. Of course this 
case has no physical basis and is employed here only to illustrate the 
effects of the scale selectivity of the dissipation on the transient 
dynamics. We note in passing that numerical simulations aiming to verify 
the dual-cascade picture (or part of this picture: either an inverse or
a direct cascade) employ both hyperviscosity and inverse viscosity. The
results in this direction constitute a rich literature; some recent studies 
are \cite{Paret99}, \cite{Boffetta00}, \cite{Lindborg00} and \cite{Chen03}. 
The present arguments suggest that the effects of these scale-selective 
dissipation mechanisms on the transient dynamics (and probably beyond) 
should be taken into consideration when interpreting the results.

For finite Reynolds numbers a small but non-negligible fraction of the
injected energy is dissipated, resulting in a weak (i.e. less than complete) 
inverse energy cascade. For the $k^{-5/3}$ energy range this dissipation 
occurs mainly around the forcing wavenumber $s$, resulting in a similar 
fraction of enstrophy dissipation around $s$. Now if the remainder of the 
enstrophy injection were to be transferred to and dissipated at $k\gg s$, 
there would necessarily be a severe step in the spectrum between the forcing 
region and the rest of the enstrophy range. This means that for power-law 
scalings (without such a step), a weak direct enstrophy cascade is not 
permitted. This is in contrast to the robustness of the weak inverse cascade, 
which is readily observable in numerical simulations even in the complete 
absence of an accompanying direct enstrophy cascade (see Tran 2004 and Tran 
\& Bowman 2004). Hence a plausible dynamical scenario is that for moderate 
Reynolds number the enstrophy dissipation occurs throughout the direct-transfer
range up to some high wavenumber, which depends on the Reynolds number. The 
energy spectrum of this enstrophy dissipation range (instead of enstrophy 
inertial range) should then scale as $k^{-5}$. The question is whether, for 
progressively higher Reynolds number, the inverse energy cascade can become 
stronger and a direct enstrophy cascade realizable. In a more quantitative 
analysis, \cite{T04} suggests that a quasi-steady state featuring an inverse 
cascade of arbitrary strength (via the Kolmogorov-Kraichnan $k^{-5/3}$ 
spectrum) and a uniform dissipation of enstrophy among the wavenumber 
octaves of the direct-transfer range is plausible. This picture is 
consistent with the preferential transfer of energy and enstrophy, required 
by the conservation laws, and explains the numerical results of \cite{T04} 
and \cite{TB04} mentioned earlier. It may also explain the numerous 
numerical results targeted at the inverse energy cascade, for which there 
is hardly an enstrophy range due to limited resolution.    

In order to apply the dual-cascade hypothesis to turbulence confined to
a bounded domain, \cite{Kraichnan67} suggests that after reaching the 
lowest available wavenumber $k_0$, the inverse energy cascade maintains 
its strength during the quasi-steady stage, depositing energy onto $k_0$
(or in the vicinity of $k_0$, as presently considered). This process 
continues until the growth of energy at or around $k_0$ is limited by its 
own dissipation, resulting in a huge pile-up of energy and enstrophy in this 
region of the spectrum. Note that even in the picture of no direct enstrophy 
cascade discussed in the preceding paragraph, the Kraichnan conjecture still 
applies although in this case the concentration of energy and enstrophy 
at the condensate would not be as dramatic as in the original dual-cascade 
case. This topic is discussed further in \S\,5.

\section{Inverse energy transfer}

In this section, we derive a rigorous upper bound for the energy that 
gets transferred across a low wavenumber $\ell$. It is shown that if 
a power-law scaling $ak^{-\alpha}$ is assumed for the energy spectrum
in the inverse-transfer range and if $\alpha<5/3$, then no significant 
fraction of the energy injection can get transferred across $\ell$ for 
sufficiently low $\ell$. This result implies that the $-5/3$ slope
represents the minimal steepness of the energy inertial range that 
can support an inverse energy cascade. For the special case $\ell=k_0$
a sharp estimate of the energy transfer onto $k_0$ is obtained.
   
\subsection{Inverse energy flux}

We assume that the spectral support of $f$ is bounded from below by a 
wavenumber $s_0$ and that the energy injection is bounded. The usual 
requirement of spectral localization of $f$ can be relaxed for most of this 
section. For $\ell<s_0$ the governing equation for the evolution of the 
large-scale energy density $E_<=\langle|\nabla\psi^<|^2\rangle/2$ is 
obtained by multiplying the governing equation (\ref{NS}) by $\psi^<$ 
and taking the spatial average of the resulting equation:
\begin{eqnarray}
\label{evolution1}
\frac{\d}{\dt}E_< &=& \langle\psi^<J(\psi,\Delta\psi)\rangle
-2\nu Z_<= -\langle\Delta\psi J(\psi,\psi^<)\rangle-2\nu Z_<,
\end{eqnarray}
where the second equality is due to (\ref{conservation}). The 
nonlinear term represents the energy that gets transferred into 
the low-wavenumber region $[k_0,\ell]$, i.e. the energy flux 
across $\ell$, which drives the large-scale dynamics. 

We now derive an upper bound for the nonlinear term in (\ref{evolution1})
and then estimate this bound for power-law energy spectra. The steps go 
as follows
\begin{eqnarray}
\label{fluxbound1}
|\langle\Delta\psi J(\psi,\psi^<)\rangle| &=&
|\langle\psi J(\psi_x,\psi^<_x)+\psi J(\psi_y,\psi^<_y)\rangle|\nonumber\\
&=& |\langle\psi J(\psi^>_x,\psi^<_x)+\psi J(\psi^>_y,\psi^<_y)\rangle| 
\nonumber\\
&=& |\langle\psi^>_x J(\psi,\psi^<_x)+\psi^>_y J(\psi,\psi^<_y)\rangle| 
\nonumber\\
&\le& \langle|\psi^>_x||\nabla\psi||\nabla\psi^<_x|+|\psi^>_y|
|\nabla\psi||\nabla\psi^<_y|\rangle \nonumber\\
&\le& \langle|\nabla\psi^>||\nabla\psi|(|\nabla\psi^<_x|^2+|\nabla\psi^<_y|^2)^
{1/2}\rangle \nonumber\\
&\le& 2E_>^{1/2}E^{1/2}\sup_{\bm x}(|\nabla\psi^<_x|^2
+|\nabla\psi^<_y|^2)^{1/2} \nonumber\\
&=& 2E_>^{1/2}E^{1/2}\sup_{\bm x}(|\psi^<_{xx}|^2+|\psi^<_{yy}|^2
+2|\psi^<_{xy}|^2)^{1/2},
\end{eqnarray}
where the first step is due to (\ref{id}) and all the subsequent steps 
involve simple manipulations and the Cauchy-Schwarz inequality. Here 
$E_>=\langle|\nabla\psi^>|^2\rangle/2$ is the `small-scale' energy density 
associated with wavenumbers $k\ge\ell$. We observe that the upper bound for 
$|\Delta\psi^<|$ in (\ref{supbound}) is also an upper bound for 
$|\psi^<_{xx}|$, $|\psi^<_{yy}|$ and $2|\psi^<_{xy}|$. Hence we can deduce that
\begin{eqnarray}
\label{fluxbound2}
|\langle\Delta\psi J(\psi,\psi^<)\rangle| &\le&  
c'\frac{\ell}{k_0}Z_<^{1/2}E_>^{1/2}E^{1/2},
\end{eqnarray}
where $c'=2\sqrt{3}c$. The appearance of $Z_<$ in (\ref{fluxbound2}) is 
due mainly to (\ref{id}), which enables us to `transfer' one spatial
derivative from $\psi$ to $\psi^<$. In the velocity formulation of the
NS system, the calculation leading to the upper bound (\ref{fluxbound2})
for the inverse flux is straightforward. One of the advantages of the 
vorticity formulation, when equipped with (\ref{id}), is that various
nonlinear terms, for example the term 
$\langle\Delta^n\psi J(\psi,\Delta\psi)\rangle$ in the evolution equation
of the quadratic quantity $\langle\psi\Delta^{n+1}\psi\rangle/2$, where
$n$ is an integer, can be manipulated with ease. Since $Z_<$ becomes 
smaller for progressively lower $\ell$, the right-hand side of 
(\ref{fluxbound2}) can remain bounded in the limit $\ell\rightarrow0$, 
even though both $E_>$ and $E$ diverge in that limit. 

For further analysis of the right-hand side of (\ref{fluxbound2}), we 
assume that an inverse-cascading range $ak^{-\alpha}$ has been established
beyond $\ell$ and possibly has reached $k_0$, but no reflection or 
accumulation of energy has yet occurred. For this spectrum, $E$, $E_>$ 
and $Z_<$ can be estimated as follows:
\begin{eqnarray}
E &\le& a\int_{k_0}^\infty k^{-\alpha}\,\dk =
\frac{a}{\alpha-1}k_0^{1-\alpha}~~~~\mbox{for}~~\alpha>1,\\
E_> &\le& a\int_{\ell}^\infty k^{-\alpha}\,\dk =
\frac{a}{\alpha-1}\ell^{1-\alpha}~~~~\mbox{for}~~\alpha>1
\end{eqnarray}
and
\begin{eqnarray}
Z_< &=& a\int_{k_0}^\ell k^{2-\alpha}\,\dk \le
\frac{a}{3-\alpha}\ell^{3-\alpha}~~~~\mbox{for}~~\alpha<3.
\end{eqnarray}
The requirement $\alpha<3$ is consistent with a persistent inverse energy 
cascade because a steeper spectrum would render the inverse-cascading range 
dissipative, which would not support such a cascade. The requirement 
$\alpha>1$ poses no loss of generality as becomes apparent shortly.
By substituting the above estimates into (\ref{fluxbound2}) we obtain   
\begin{eqnarray}
\label{fluxbound3}
|\langle\Delta\psi J(\psi,\psi^<)\rangle| &\le&  
\frac{c'a^{3/2}}{(\alpha-1)(3-\alpha)^{1/2}}\left(\frac{\ell}{k_0}\right)^
{(\alpha+1)/2}\ell^{(5-3\alpha)/2}.
\end{eqnarray}
Now in the limit $k_0\rightarrow0$, we have 
\begin{eqnarray}
\label{zbound}
Z &\ge& a\int_0^{s_0}k^{2-\alpha}\,\dk = \frac{a}{3-\alpha}s_0^{3-\alpha}.
\end{eqnarray}
Solving for $a$ from (\ref{zbound}) and substituting the result into
(\ref{fluxbound3}) we obtain
\begin{eqnarray}
\label{fluxbound4}
|\langle\Delta\psi J(\psi,\psi^<)\rangle| &\le&  
\frac{c'(3-\alpha)Z^{3/2}}{(\alpha-1)s_0^2}\left(\frac{\ell}{k_0}\right)^
{(\alpha+1)/2}\left(\frac{\ell}{s_0}\right)^{(5-3\alpha)/2}.
\end{eqnarray}
An interesting conclusion can be readily drawn from (\ref{fluxbound4}).
Given a fixed ratio $\ell/k_0$ and finite enstrophy density $Z$ (recall 
that we must have $Z \leq \epsilon/2\nu$), if $\alpha<5/3$, then the 
right-hand side of (\ref{fluxbound4}) can be made arbitrarily 
small provided $\ell/s_0$ is sufficiently small. In other words, no energy 
inertial range shallower than $k^{-5/3}$ would be capable of sustaining 
an inverse energy cascade that carries a nonzero fraction of the energy 
injection to the low-wavenumber region $[k_0,\ell]$, for sufficiently low
$\ell$. Thus the Kolmogorov--Kraichnan $k^{-5/3}$ spectrum represents the 
shallowest possible spectrum that is necessary for the existence of the
classical inverse energy cascade.

\subsection{Energy transfer onto $k_0$}

When $\ell=k_0$, the upper bound for the inverse flux derived in the 
preceding subsection becomes an upper bound for the energy that gets 
transferred onto $k_0$. For this special case, some improvement on 
the bound is possible and a sharp estimate for the Kolmogorov constant
can be derived. 
 
We employ the notation $\psi=\psi^<+\psi^>$ as in the previous sections, 
where $\ell=k_0$. That means $\psi^<$, which is replaced by $\psi^0$ in 
what follows, consists of only four degenerate components corresponding 
to the four wavevectors $(\pm k_0,0)$ and $(0,\pm k_0)$. Namely,
\begin{eqnarray}
\psi^0=\varphi_1\exp\{ik_0x\} + \varphi_1^*\exp\{-ik_0x\} 
+ \varphi_2\exp\{ik_0y\} + \varphi_2^*\exp\{-ik_0y\}. 
\end{eqnarray} 
Inequality (\ref{supbound}) becomes
\begin{eqnarray}
\label{estimate}
|\Delta\psi^0| &\le& 2k_0^2(|\varphi_1|+|\varphi_2|)
\le 2^{3/2}k_0\left(|k_0^2\varphi_1|^2+k_0^2|\varphi_2|^2\right)^{1/2}
= 2^{3/2}k_0\Psi_2^{1/2}(k_0),  
\end{eqnarray}  
where $\Psi_2(k_0)=\langle|\nabla\psi^0|^2\rangle/2$ is the modal energy
associated with $k_0$. 

The energy that gets transferred onto $k_0$, i.e. the nonlinear term
$\langle\psi^0J(\psi,\Delta\psi)\rangle$, can be estimated as follows: 
\begin{eqnarray}
\label{fluxbound5}
|\langle\psi^0J(\psi,\Delta\psi)\rangle| &=&
|\langle\psi^0J(\psi^>,\Delta\psi^>)\rangle| 
= |\langle\Delta\psi^> J(\psi^>,\psi^0)\rangle| \nonumber\\
&=& |\langle\psi^> J(\psi^>_x,\psi^0_x)\rangle
+\langle\psi^> J(\psi^>_y,\psi^0_y)\rangle| \nonumber\\
&=& |\langle\psi^>_x J(\psi^>,\psi^0_x)\rangle
+ \langle\psi^>_y J(\psi^>,\psi^0_y)\rangle| \nonumber\\
&=& |\langle\psi^>_x\psi^>_y(\psi^0_{yy}-\psi^0_{xx})\rangle|
\le 2k_0^2(|\varphi_1|+|\varphi_2|)\langle|\psi^>_x\psi^>_y|\rangle 
\nonumber\\
&\le& 2^{3/2}k_0\Psi_2^{1/2}(k_0)E_>,
\end{eqnarray}
where the replacement of $\psi$ by $\psi^>=\psi-\psi^0$ in the second
step is a consequence of both $\Delta\psi^0=-k_0^2\psi^0$ and 
(\ref{conservation}), the third step is due to (\ref{id}), the last 
step is due to (\ref{estimate}) and all other steps are straightforward. 
Here $E_>=\langle|\nabla\psi^>|^2\rangle/2$ is the total energy 
density with the contribution from $k_0$ removed. The nonlinear term in 
the last equation of (\ref{fluxbound5}) represents an upper bound on 
the energy that gets transferred onto $k_0$.
 
As in the previous case we assume an inverse-transfer range $ak^{-\alpha}$ 
down to $k_0$ and estimate the energy that gets transferred onto $k_0$, 
before the arrest of the inverse cascade would deform the assumed power-law 
spectrum. The factors of the nonlinear term in (\ref{fluxbound5}) can be 
estimated as follows:
\begin{eqnarray}
E_> &\le& a\int_{\sqrt{2}k_0}^\infty k^{-\alpha}\,\dk =
\frac{2^{(1-\alpha)/2}a}{\alpha-1}k_0^{1-\alpha},\nonumber\\
\Psi_2(k_0) &=& \frac{k_0E(k_0)}{2\pi} = \frac{a}{2\pi}k_0^{1-\alpha},
\end{eqnarray}
where $2^{1/2}k_0$ is the second lowest wavenumber, and 
$E(k_0)=ak_0^{-\alpha}$ is the energy supported at $k_0$. It follows that
\begin{eqnarray}
\label{fluxbound6}
2^{3/2}k_0\Psi_2^{1/2}(k_0)E_> &\le&
\frac{2^{1-\alpha/2}a^{3/2}}{\pi(\alpha-1)}k_0^{(5-3\alpha)/2}.
\end{eqnarray}
Like the argument in the preceding subsection, ineq. (\ref{fluxbound6}) 
together with (\ref{fluxbound5}) implies that an inverse-cascading range 
shallower than $k^{-5/3}$ is incapable of supporting a nonzero transfer 
of energy to $k_0$ for sufficiently low $k_0$. 

\subsection{Bounds for the Kolmogorov constant}
Suppose that the turbulence is driven by a steady energy injection rate
$\epsilon$ and that an inverse cascade carrying a fraction $r$ of this 
injection toward $k_0$ via a $ak^{-5/3}$ energy inertial range has been 
established. The left-hand side of (\ref{fluxbound6}), which is an upper 
bound for the inverse energy transfer onto $k_0$, cannot be smaller than 
$r\epsilon$. Hence we can deduce from (\ref{fluxbound6}) that
\begin{eqnarray}
\label{a}
a &\ge& \left(\frac{2^{5/6}\pi r\epsilon}{3}\right)^{2/3}.
\end{eqnarray} 
It follows that the Kolmogorov constant $C$ which appears in the energy
spectrum of the classical energy inertial range as 
$E(k)=C\epsilon^{2/3}k^{-5/3}$ is bounded from below by 
\begin{eqnarray}
\label{C}
C &\ge& \left(\frac{2^{5/6}\pi r}{3}\right)^{2/3}.
\end{eqnarray}
Note that in the classical case $r=1$. 

An upper bound for $C$ can be derived on the basis of the classical
spectrum alone. We observe that the energy dissipation by the energy 
range cannot exceed $(1-r)\epsilon$. Hence we have
\begin{eqnarray}
2\nu C\epsilon^{2/3}\int_0^sk^{1/3}\,\dk &\le& (1-r)\epsilon.
\end{eqnarray}
It follows that
\begin{eqnarray}
\label{C1}
C &\le& \frac{2(1-r)\epsilon^{1/3}}{3\nu s^{4/3}}.
\end{eqnarray}
If we assume that $\epsilon$ is injected around a forcing wavenumber $s$,
so that the enstrophy injection is given by $\eta=s^2\epsilon$, then we
can rewrite (\ref{C1}) as 
\begin{eqnarray}
\label{C2}
C &\le& \frac{4(1-r)\eta^{1/3}}{3\tau},
\end{eqnarray}
where $\tau=2\nu s^2$ is the dissipation rate at the forcing wavenumber 
$s$. The upper bound (\ref{C2}) is expected to be sharp in the limit of 
high Reynolds number. In fact (\ref{C2}) is essentially an equality in 
that limit since the energy dissipation by $k>s$ is negligible. For a 
fixed enstrophy injection rate $\eta$, constancy of $C$ requires that the 
ratio $(1-r)/\tau$ remain constant. On the other hand, $C$ may diverge 
in the limit of infinite Reynolds number ($\tau\rightarrow0$) provided 
the inverse-cascade strength $r$ approaches unity slower than
$\tau$ tends to zero. This dynamical scenario, which has been touched upon in
\S\,3, is considered by \cite{T04} for the unbounded case, where it is 
suggested that an inverse energy cascade that is progressively stronger 
with progressively higher Reynolds number is quite plausible, requiring 
no boundedness of enstrophy or of any other quadratic quantities in the 
limit of infinite Reynolds number. In this picture there is no anomalous 
enstrophy dissipation: the enstrophy dissipation is uniformly distributed 
among the wavenumber octaves of the direct-transfer range, whose energy 
spectrum scales as $k^{-5}$.
 
\section{Power-law spectra vs. Kraichnan's condensate}

In this section we examine the Kraichnan picture for steady dynamics in a 
bounded domain, and 
derive constraints on the enstrophy and palinstrophy distribution for such 
a picture. We then contrast this picture with power-law spectra. For 
simplicity we follow the non-average treatment in the preceding sections.
Nevertheless all dynamical quantities, including the energy spectrum, can
be understood in an appropriate average sense since such a reinterpretation
requires a straightforward reformulation of the problem. 

By virtue of the conservation laws, the evolution of the energy and 
enstrophy is governed by
\begin{eqnarray}
\label{governing}
\frac{\d E}{\dt} &=& -2\nu Z + \epsilon,\\
\frac{\d Z}{\dt} &=& -2\nu P + \eta,
\end{eqnarray}
where $\epsilon=-\langle f\psi\rangle$ and $\eta=\langle f\Delta\psi\rangle$
are, respectively, the energy and enstrophy injection rates. The usual 
assumption of spectral localization of $f$ around $s$ is invoked, i.e. 
$\eta=s^2\epsilon$. The forced-dissipative balance of both energy and 
enstrophy implies (see Constantin \etal 1994, Eyink 1996, Tran \& Shepherd 
2002 and Kuksin 2004) 
\begin{eqnarray}
\label{balance}
P = s^2 Z, 
\end{eqnarray}
or in terms of the energy spectrum $E(k)$,
\begin{eqnarray}
\label{balance1}
\int_{k_0}^\infty(k^2-s^2)k^2E(k)\,\dk = 0.
\end{eqnarray} 
Equation (\ref{balance1}) is the focus of the present section.

\subsection{Energy and enstrophy condensate}

Eq. (\ref{balance}), which is independent of Reynolds  number, has been 
interpreted as implying that the enstrophy dissipation must occur in the 
vicinity of $s$, thus precluding a direct entrophy cascade. However, this 
interpretation assumes that all the enstrophy is actively involved in the 
enstrophy cascade. As noted earlier, the Kraichnan energy condensate is 
also an enstrophy condensate, in which case most of the enstrophy is 
trapped in the condensate and only a tiny amount is free to participate 
in the enstrophy cascade. Thus, this interpretation implicitly excludes 
the Kraichnan scenario.

In fact, for spectra such that $Z$ is distributed mainly around $k_0$ and 
$P$ is distributed mainly at $k\gg s$, the balance (\ref{balance}) 
can still hold (see below). Another example is that $P$ has an equal 
contribution from each of the wavenumber octaves 
$[s,10s],~[10s,10^2s],\cdots$, up to some wavenumber 
$k_\nu\approx s^2/k_0$, provided that $Z$ has a similar contribution 
from the wavenumber octaves of $[k_0,s]$. This uniform distribution 
of the energy and enstrophy dissipation in their respective ranges 
(requiring $k^{-3}$ energy and $k^{-5}$ enstrophy ranges) is considered 
by \cite{TB03} and \cite{T04} (also see below). The former example 
includes the Kraichnan scenario, for which the energy of the condensate 
$E_<$ (with $\ell\approx k_0$) satisfies $E_<\approx\epsilon/2\nu k_0^2$. 
The energy dissipation occurs mainly at the condensate since 
$2\nu Z_<\approx\epsilon$, where $Z_<\approx k_0^2E_<$ is the enstrophy 
of the condensate. The enstrophy dissipation by the condensate is 
$2\nu P_<\approx k_0^2\epsilon$, where $P_<\approx k_0^2Z_<$ is the 
palinstrophy of the condensate. This dissipation is negligible as
compared with the enstrophy injection $s^2\epsilon$ if $k_0/s\ll1$, i.e. 
if the energy inertial range is sufficiently wide. This possibility 
allows for the spectral distribution of palinstrophy, i.e. the spectral 
distribution of enstrophy dissipation, to be concentrated at the high 
wavenumbers, while posing no threat to the balance $P=s^2Z$. 

The discussion in the preceding paragraph implies that the Kraichnan
energy condensate would also be an enstrophy condensate. The extent of 
this highly localized enstrophy distribution at the condensate would be
comparable to that of the palinstrophy around the dissipation wavenumber 
$k_\nu$. To see this explicitly let us rewrite the balance equation
(\ref{balance}) as
\begin{eqnarray}
P_<+P_> = s^2(Z_<+Z_>),
\end{eqnarray}
where the subscripts `$<$' and `$>$' refer to the `trapped' (in the 
condensate) and `free' quantities. It follows that
\begin{eqnarray}
\frac{P_>}{s^2Z_>} = 1+\left(1-\frac{k_c^2}{s^2}\right) 
\frac{Z_<}{Z_>}, 
\end{eqnarray}
where $k_c^2=P_</Z_<\approx k_0^2$. In the Kraichnan picture the ratio 
$P_>/s^2Z_>\approx k_\nu^2/s^2$ diverges in the limit of infinite Reynolds 
number. Hence the ratio $Z_</Z_>$ diverges in a similar manner. Thus the 
Kraichnan energy condensate is also an enstrophy condensate. 

\subsection{Power-law spectra}

Following Constantin \etal (1994), \cite{T04} and \cite{TB03}, 
we assume a two-range energy spectrum:
\begin{eqnarray}
\label{spectrum}
E(k) &=& \cases{
ak^{-\alpha}&if $k_0 < k < s$,\cr
bk^{-\beta}&if $s < k < k_\nu$,\cr} ~~~ as^{-\alpha} = bs^{-\beta},
\end{eqnarray} 
where $\alpha$ and $\beta$ are constant and $k_\nu$ marks the end of 
the $k^{-\beta}$ range, beyond which the spectrum becomes steeper. 
This approximation a priori excludes the Kraichnan condensate, but 
allows for some `milder' accumulation of energy at $k_0$. For example, 
the case $\alpha=3$ previously considered would allow for most of the 
system's energy to reside in a few lowest wavenumbers. Substituting 
(\ref{spectrum}) into (\ref{balance1}) allows one to derive interesting 
constraints on $\alpha$ and $\beta$. By rewriting (\ref{balance1}) in 
the form 
$\int_{k_0}^s(s^2-k^2)k^2E(k)\,\dk=\int_s^\infty(k^2-s^2)k^2E(k)\,\dk$,
using the approximation (\ref{spectrum}) for $E(k)$ and making the 
substitutions $\kappa=k/s$ for $k\le s$, and $\kappa=s/k$ for $k\ge s$, 
\cite{TB03} show that
\begin{eqnarray}
\label{constraint}
\int_{k_0/s}^1(1-\kappa^2)\kappa^{2-\alpha}\,\d\kappa \ge
\int_{s/k_\nu}^1(1-\kappa^2)\kappa^{\beta-6}\,\d\kappa,
\end{eqnarray} 
where the inequality is a consequence of dropping from the equality
the contribution beyond $k_\nu$, i.e. the quantity 
$D_\nu=\int_{k_\nu}^\infty(k^2-s^2)k^2E(k)\,\dk$. Except for the 
approximation (\ref{spectrum}), the constraint (\ref{constraint}) is 
rigorous. In the spirit of KLB ($k_\nu\rightarrow\infty$ as 
$\nu\rightarrow0$), we consider the case $k_0/s\ge s/k_\nu$, which 
corresponds to high-Reynolds number turbulence. It is then easy to 
see from (\ref{constraint}) that $2-\alpha\le\beta-6$, or equivalently,
\begin{eqnarray}
\label{constraint1}
\alpha+\beta \ge 8.
\end{eqnarray}
For moderate Reynolds number $\beta>5$ (see Tran 2004 and Tran \& Bowman
2004), so we first consider the case $\beta\ge5$. The quantity $D_\nu$ 
is then negligible, making (\ref{constraint}) essentially an equality and 
(\ref{constraint1}) essentially an equality if $k_0/s\approx s/k_\nu$. 
In this case we obtain $(\alpha,\beta)=(3-\delta,5+\delta)$, where the 
limit $\delta\rightarrow0$ is expected for high Reynolds numbers. This 
solution with $\delta=0$ means that the energy and enstrophy are 
uniformly dissipated among the wavenumber octaves of their respective 
ranges. On the other hand, if $\beta<5$ (which cannot be logically 
ruled out, just as the Kraichnan picture cannot be a priori excluded), 
then $\alpha>3$. In this case the inequality ``$\ge$'' in 
(\ref{constraint}) could become ``$\gg$'' since $D_\nu$ could become 
large. As a consequence a slight decrease of $\beta$ from the critical 
value $\beta=5$ must be met with a much more significant increase of 
$\alpha$ from $\alpha=3$. This means that $\alpha$ could quickly approach 
$\beta$ before the latter would drop significantly below its critical
value. Hence the two-range steady spectrum (\ref{spectrum}), when 
supplemented by the usual condition $\alpha<\beta$, obeys the constraint 
$\beta\ge 4+\delta$, where $\delta>0$ depends on how $E(k)$ decays for 
$k>k_\nu$. The case $\beta\le4$ is inadmissible and would necessarily 
require an energy condensate. For the Kolmogorov--Kraichnan 
energy inertial range ($\alpha=5/3$), a Kraichnan-type condensate would 
also be required even when $\beta=5$ since the integral on the left-hand 
side of (\ref{constraint}) is less than unity (for all ratios $k_0/s$)
while its counterpart on the right-hand side is $\approx\ln(k_\nu/s)$,
which diverges as $k_\nu/s\rightarrow\infty$. Finally, the classical 
$k^{-3}$ direct-cascade range requires a $k^{-5}$ inverse-cascade range
if only power-law spectra are permitted. The enstrophy content of the 
Kraichnan condensate would be equivalent to the enstrophy content of 
a $k^{-5}$ energy range.

\section{Conclusion and discussion}

In this paper the inverse energy transfer in 2D NS turbulence is studied.
The main result obtained is an upper bound for the inverse energy flux.
This upper bound is then estimated using power-law spectra. It is shown 
that a steady inverse energy flux requires that the energy spectrum of the
inverse-transfer range be at least as steep as $k^{-5/3}$: a shallower 
spectrum is unable to support an inverse energy cascade, and the inverse 
energy flux necessarily diminishes as it proceeds toward sufficiently low 
wavenumbers. This result provides for the first time a rigorous basis for 
the Kolmogorov-Kraichnan $k^{-5/3}$ energy inertial range in 2D turbulence. 
Other results include estimates of the Kolmogorov constant, a relation 
between the enstrophy and palinstrophy distribution for the Kraichnan 
condensate, and an analysis of power-law scaling in contrast to the
Kraichnan conjecture.

The inverse energy flux across a wavenumber $\ell$ in the energy range is 
found to depend on the system's energy and on the enstrophy content of 
the wavenumber region lower than $\ell$. The former is supposed to grow
without bound in the limit of unbounded domain, while the latter becomes
progressively smaller for progressively lower $\ell$. These two effects 
counterbalance each other in such a way that the classical inverse energy 
cascade necessarily requires that the energy spectrum of the inverse-transfer 
range be at least as steep as the Kolmogorov--Kraichnan $k^{-5/3}$ spectrum.

The derived upper bound for the inverse energy flux has been used to
deduce a lower bound for the Kolmogorov constant $C$ in the classical
energy spectrum $E(k)=C\epsilon^{2/3}k^{-5/3}$. This bound is of order
unity and is expected to hold whether or not $C$ is a universal constant. 

We have elaborated on the Kraichnan conjecture of energy condensation. 
A constraint on the enstrophy and palinstrophy distribution for this 
picture has been derived and contrasted with those for power-law scalings.
For power-law spectra, a $k^{-3}$ energy range and a $k^{-5}$ enstrophy
range are consistent with the forced-dissipative balance of energy and 
enstrophy. This new picture has some justification from the numerical 
results of \cite{Borue94}, \cite{T04} and \cite{TB04}, at least for 
moderate Reynolds numbers. A Kraichnan-type condensate is required for 
energy-range spectra shallower than $k^{-3}$ even if the enstrophy-range 
spectrum remains as steep as $k^{-5}$. The energy and enstrophy level 
of the condensate would dramatically increase should both the energy
range be shallower than $k^{-3}$ and the enstrophy range be shallower 
than $k^{-5}$. The classical $k^{-3}$ enstrophy range would require an
excessively high concentration of energy and enstrophy at the condensate:
the enstrophy content of the condensate would be equivalent to that
of a $k^{-5}$ energy range.

Several authors have considered steady bounded turbulence and, on the 
basis of the balance equation (\ref{balance}), ruled out the existence of a 
direct enstrophy cascade (see Constantin \etal 1994; Eyink 1996, p. 110; 
Tran \& Shepherd 2002; Kuksin 2004). These works either explicitly or 
implicitly exclude a priori the Kraichnan conjecture in favour of 
power-law spectra. In particular Constantin \etal (1994) assume the
power-law scaling described by (\ref{spectrum}) with the classical 
exponents $\alpha=5/3$ and $\beta=3$, and find that this spectrum is
incompatible with the balance $P=s^2Z$. However, the two-range spectrum 
of KLB is supposed to be only quasisteady, with the inverse energy cascade 
proceeding toward ever-lower wavenumbers, carrying with it virtually all 
the energy injection. In other words, the KLB spectrum is supposed to 
correspond to $\d E/\dt=-2\nu Z+\epsilon\approx\epsilon$, or equivalently, 
to $P\gg s^2 Z$, not to the balance $P=s^2Z$, which is achieved only in 
a steady state. The quasisteady limit as $t \to \infty$ is clearly 
incompatible with a finite domain. \cite{Eyink96} and \cite{TS02} infer 
from (\ref{balance}) that the dissipation of enstrophy occurs mainly in 
the vicinity of the wavenumber given by $\sqrt{P/Z}$, which is just the 
forcing wavenumber $s$ for steady turbulence, thereby ruling out the 
existence of a direct enstrophy cascade. This conclusion is valid for 
moderate ratios $s/k_0$, i.e. for relatively narrow energy ranges, and 
for various `regular' spectra such as ones with the energy range shallower 
than $k^{-3}$. However if the Kraichnan conjecture holds, then this 
conclusion would not be valid. \cite{Kuksin04} 
employs a special forcing, such that both the energy and enstrophy 
injections are proportional to the viscosity coefficient $\nu$,
to show that in the limit of infinite Reynolds number the statistical 
equilibrium palinstrophy remains bounded. This result leads Kuksin to 
conclude that the KLB theory does not apply to the statistically steady 
dynamics of bounded NS turbulence in a doubly periodic domain. Similar
to \cite{Eyink96} and \cite{TS02}, the balance $P=s^2Z$ is also obtained,
but no further constraints related to the Kraichnan scenario are derived. 
The fact that $P$ remains bounded in the limit of infinite Reynolds number 
is due to the special choice of the forcing. In essence this result is
equivalent to that of \cite{Eyink96} and \cite{TS02}, except by a scaling 
factor of the inverse of the Reynolds number. Hence the above analyses
apply to this case as well. In particular the Kraichnan conjecture of 
energy condensation cannot be logically excluded. The present work 
emphasizes the need for some treatment beyond the forced-dissipative 
balances of the energy and enstrophy, before such an exclusion can be 
fully justified.     

\begin{acknowledgments} 
This work was motivated by the discussions at the American Institute 
of Mathematical Sciences' Fifth International Conference on Dynamical 
Systems and Differential Equations. We would like to thank Gregory 
Eyink for pointing out the difficulty in the interpretation of the 
balance constraint, in light of the Kraichnan conjecture. We would also 
like to thank Ka-Kit Tung for organizing a highly successful session on 
2D and QG turbulence. Helpful discussions with J. Tribbia, M. Jolly, 
J. Bowman, J. Mattingly, P. Fischer, and E. Gkioulekas are gratefully 
acknowledged.
\end{acknowledgments}

\end{document}